\documentclass[twocolumn,trackchanges]{aastex631}
\usepackage{graphicx}
\usepackage{amsmath}
\usepackage{lipsum} 

\newenvironment{sizeddisplay}[1]
 {\par\nopagebreak#1\noindent\ignorespaces}
 {\nopagebreak\ignorespacesafterend}

\begin{document}

\turnoffeditone

\title{Detecting and Classifying Flares in High-Resolution Solar Spectra\\ with Supervised Machine Learning}

\author[0009-0009-8530-6183]{Nicole Hao}
\affiliation{Cornell University, Ithaca, NY 14853, USA}

\author[0000-0001-6362-0571]{Laura Flagg}
\affiliation{Department of Physics and Astronomy, Johns Hopkins University, 3400 N. Charles Street, Baltimore, MD 21218, USA}
\affiliation{Cornell University, Ithaca, NY 14853, USA}
\correspondingauthor{Laura Flagg}
\email{laura.s.flagg@gmail.com}

\author[0000-0001-5349-6853]{Ray Jayawardhana}
\affiliation{Department of Physics and Astronomy, Johns Hopkins University, 3400 N. Charles Street, Baltimore, MD 21218, USA}



\begin{abstract}

Flares are a well-studied aspect of the Sun's magnetic activity. Detecting and classifying solar flares can inform the analysis of contamination caused by stellar flares in exoplanet transmission spectra. In this paper, we present a standardized procedure to classify solar flares with the aid of supervised machine learning. Using flare data from the RHESSI mission and solar spectra from the HARPS-N instrument, we trained several supervised machine learning models, and found that the best performing algorithm is a C-Support Vector Machine (SVC) with non-linear kernels, specifically Radial Basis Functions (RBF). The best-trained model, SVC with RBF kernels, achieves an average aggregate accuracy score of 0.65, and categorical accuracy scores of over 0.70 for the no-flare and weak-flare classes, respectively. In comparison, a blind classification algorithm would have an accuracy score of 0.33.  Testing showed that the model is able to detect and classify solar flares in entirely new data with different characteristics and distributions from those of the training set. Future efforts could focus on enhancing classification accuracy, investigating the efficacy of alternative models, particularly deep learning models, and incorporating more datasets to extend the application of this framework to stars that host exoplanets.\\

\end{abstract}



\section{Introduction} \label{sec:intro}
Transmission spectroscopy is highly useful and widely used for characterizing exoplanets since it can yield valuable constraints on the nature and composition of planetary atmospheres. Yet, due to the inhomogeneity and time variability of the stellar photo- and chromospheres, this method is intrinsically impacted by stellar spectral contamination \citep{rackham_effect_2023}. Often, the stellar contamination will rival or even exceed the planetary spectral features, making it very difficult to disentangle the exoplanet atmospheric signals from stellar contamination \citep{rackham_effect_2023}. 

\edit1{Stellar flares are another source of stellar contamination, resulting in increased brightness in exoplanet transmission spectra. They are wavelength-dependent, which leads to additional noise and distortions of the observed spectra across a range of wavelengths.} Such contamination poses a challenge for measuring an exoplanet's transit depth accurately. To consider the impact of stellar flares, it is helpful to begin with an investigation of solar flares, given the abundance of flare data for the Sun. Efficient detection and classification methods for solar flares in transmission spectra could help astronomers correct for stellar contamination in exoplanet transmission spectra with greater accuracy.

This study has been designed with two primary objectives. First, it aims to address the challenge of detecting flare events in high-resolution solar spectra. To achieve this, we correlated solar spectra with solar flare events based on their start and end times, labeled solar spectra, and fed labeled solar spectra into supervised machine-learning models, which enabled accurate detection of flares based on their energy levels. Second, the project strives to develop a robust machine-learning model that can classify flares in solar spectra as accurately as possible. Identifying low-energy flares in high-resolution solar spectra may present a greater difficulty compared to their high-energy counterparts. However, the impact of low-energy flares on spectra should not be disregarded. Thus, we aim to detect and classify all flares in solar spectra, regardless of their energy levels. 

In this study, we employed supervised learning algorithms, specifically Support Vector Classification (SVC), to detect and classify solar flares in high-resolution solar spectra. Our methodology encompasses a standardized procedure for classifying flares, primarily based on their energy levels. To assess the performance of the classification models, we utilized multi-label metrics such as accuracy scores. By leveraging these evaluation measures, we aim to provide a framework for detecting and categorizing solar flares in the context of high-resolution spectra.

The paper is organized as follows. Section 2 considers the datasets used, followed in section 3 by a description of our methods, including data selection, analysis and model testing. Section 4 addresses the selection of supervised learning algorithms and the validation of results using multi-label metrics, and includes a discussion on the optimization of model performance through means such as performing grid-search and tweaking hyper-parameters. 

\section{Data} 
\begin{figure}[ht]
\centering
\includegraphics[scale = 0.43]{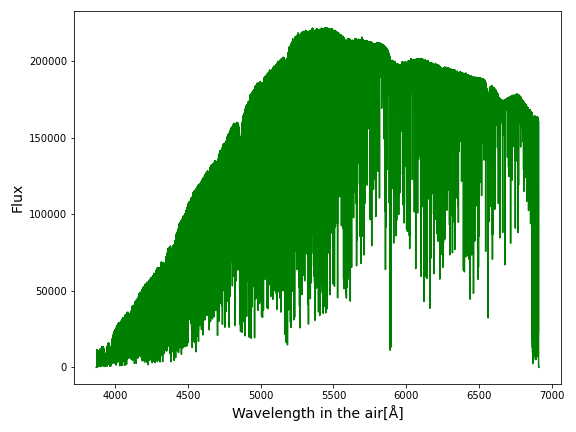}
\caption{Solar spectrum, a single observation}
\label{fig:spectra}
\end{figure}

Solar spectra, ranging from July 2015 to April 2018, were drawn from the Data \& Analysis Center for Exoplanets \citep{data__analysis_center_for_exoplanets_dace_solar_nodate}. 
The spectra were collected with HARPS-N, an optical spectrograph installed at the Italian Telescopio Nazionale Galileo (TNG), designed specifically for the precise measurement of radial velocities in the search for exoplanets \citep{cosentino_harps-n_2012}. The instrument offers a high resolution of $\sim$120,000, allowing for detailed spectral analysis, and covers a range of wavelengths \edit1{378 nm -- 691 nm}. Figure \ref{fig:spectra} shows the flux from one observation against wavelength in the air. One observation here refers to the single set of data collected by the HARPS-N over 5 minutes, capturing the intensity of light across different wavelengths from the Sun \citep{dumusque_three_2021}.

Solar flare data used in this investigation were collected by the Reuven Ramaty High Energy Solar Spectroscopic Imager (RHESSI) \citep{nasa_goddard_space_flight_center_rhessi_2003}. RHESSI was a NASA satellite mission designed to study the Sun at high-energy X-ray and gamma-ray wavelengths, with a focus on investigating the particle acceleration and energy release processes in solar flares \citep{lin_reuven_2003}. The RHESSI flare list covers a time period of approximately 17 years from 2002 to 2019. Since our study includes solar spectra only for the \edit1{2015 -- 2018} period, we used solar flare data for the same date range. The RHESSI data contain the following information for each observation: start times and end times, solar peak times, energy levels, duration, total peak counts, X-position and Y-position of the event on the solar disk, radial distance of a solar flare event from the center of the Sun, active region, and flags. We removed data points from the dataset that contained zero information. For example, if the X-position and Y-position of one observation are both zero, the Spectroscopic Imager on the satellite failed to collect any information, therefore the data point has to be removed to prevent it from introducing errors into the machine learning results. \\

\section{Methods} 
\subsection{Selection of Wavelength Range, Normalization, and Principal Component Analysis}
For this study, we used solar spectra spanning \edit1{6400 {\AA} -- 6700 {\AA}} as our training data. We selected this particular wavelength range because it contains H$\alpha$, which exhibits increased emission during a solar flare \citep{ichimoto_h_1984}.

We conducted a series of pre-processing steps on the data: First, we plotted the solar spectra for all distinct observations and identified the region exhibiting the least fluctuation in flux. Next, we calculated the 98th percentile within that region and divided the flux values of all data points by this value, generating a set of normalized flux values. 

\begin{figure}[ht]
    \includegraphics[scale=0.4]{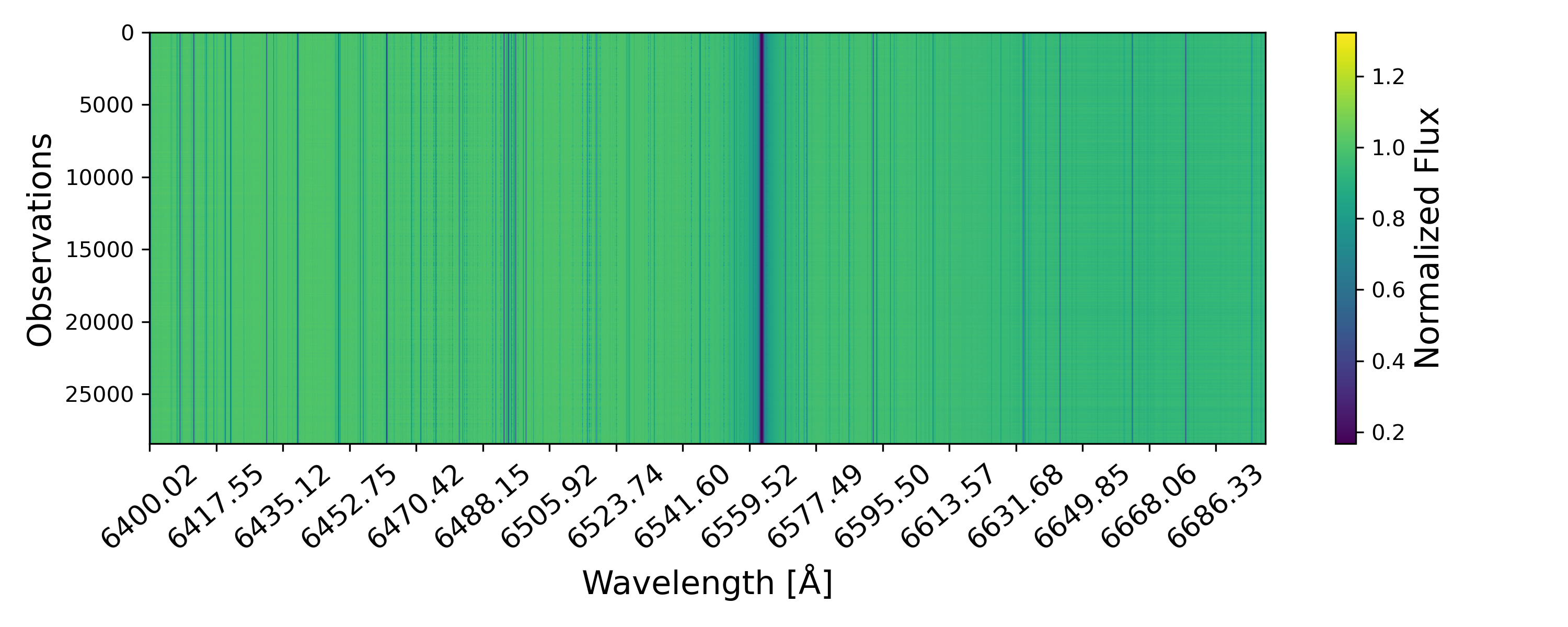}
    \caption{Plot of normalized flux against wavelengths}
    \label{fig:normflux}
\end{figure}

\begin{figure}[ht]
    \centering
    \includegraphics[scale= 0.43]{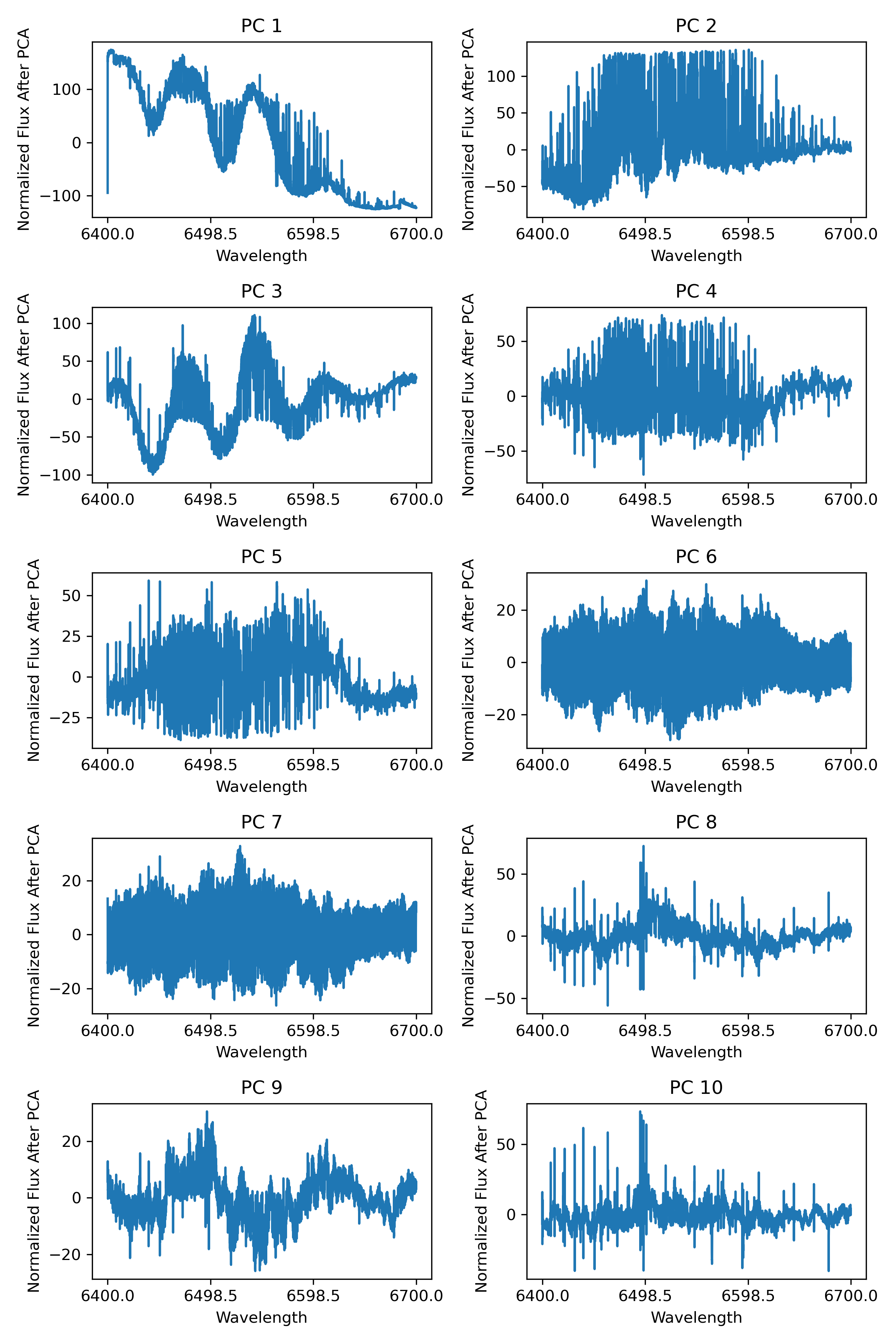}
    \caption{First 10 Principal Components of Normalized Solar Spectra}
    \label{fig:PCAplot}
\end{figure}

To reduce the effects of high-noise data on our training results, specifically outliers that appear as spikes on the plot of a solar spectrum observation, we replaced all of the normalized flux values that are five standard deviations or more from the mean of the normalized spectra with the average of their neighboring values. See figure \ref{fig:normflux} for a plot of noise-filtered normalized spectra. Then we applied Principal Component Analysis (PCA) to the processed data. PCA is a mathematical technique that reduces the dimensionality of training data by projecting data points onto a lower dimensional space that best captures the variance of the original data, therefore enhancing the efficiency of the subsequent machine learning process \citep{abdi_principal_2010}. Our processed data initially consisted of 28,415 observations and 16,747 wavelength bins. After PCA, we reduced the dimension of the normalized data to (28415, 1000) with 1000 principal components. Figure \ref{fig:PCAplot} shows the first 10 principal components of the normalized solar spectra.

\subsection{Data Correlation and Labeling}
Supervised learning algorithms take in a collection of labeled data. The labeled solar spectra represent the input values and their labels represent the output values. We approximated the true relationship between the input value and their labels using machine learning by labeling each normalized solar spectrum observation after PCA as one of the following three labels: no flares, weak flares, strong flares. We extracted flare events from the RHESSI data and their corresponding energy band. In the RHESSI data, each observation of a potential flare is considered a ``flare event
", and each has a corresponding range of energy that was observed in the duration of the flare, referred to as the event's ``energy band". The flare events were divided into three label classes: spectra without any flare events 
are categorized as ``no flares"; events with the energy band \edit1{6 -- 12 keV} are categorized as ``weak flares"; All solar events with energy bands greater than \edit1{6 -- 12 keV} are categorized as ``strong flares".  Visually, there are no differences between the spectra despite some spectra having flares and others not.

    \label{fig:strongweak}

\edit1{The process of data correlation and labeling must also consider the temporal differences between the observations from RHESSI and Halpha line data. RHESSI operates in the high-energy regime from soft X-rays to gamma rays within the coronal region, whereas the Halpha line originates from the chromosphere. Despite documented temporal differences between these regions, as supported by relevant literature, these differences are not consistently predictable \citep{gontikakis_transition_2023}. Given the variability and our lack of precise knowledge about these time discrepancies, our investigation does not explicitly account for them. Additionally, solar flares, which are the focus of our study, persist over extended periods rather than occurring as instantaneous events. Our approach assumes the effectiveness of our current method, and the performance of our final machine learning model supports this assumption, showing satisfactory results.} 

\subsection{Imbalanced Data and Under-sampling}

\begin{figure}[ht]
\centering
\includegraphics[scale=0.2]{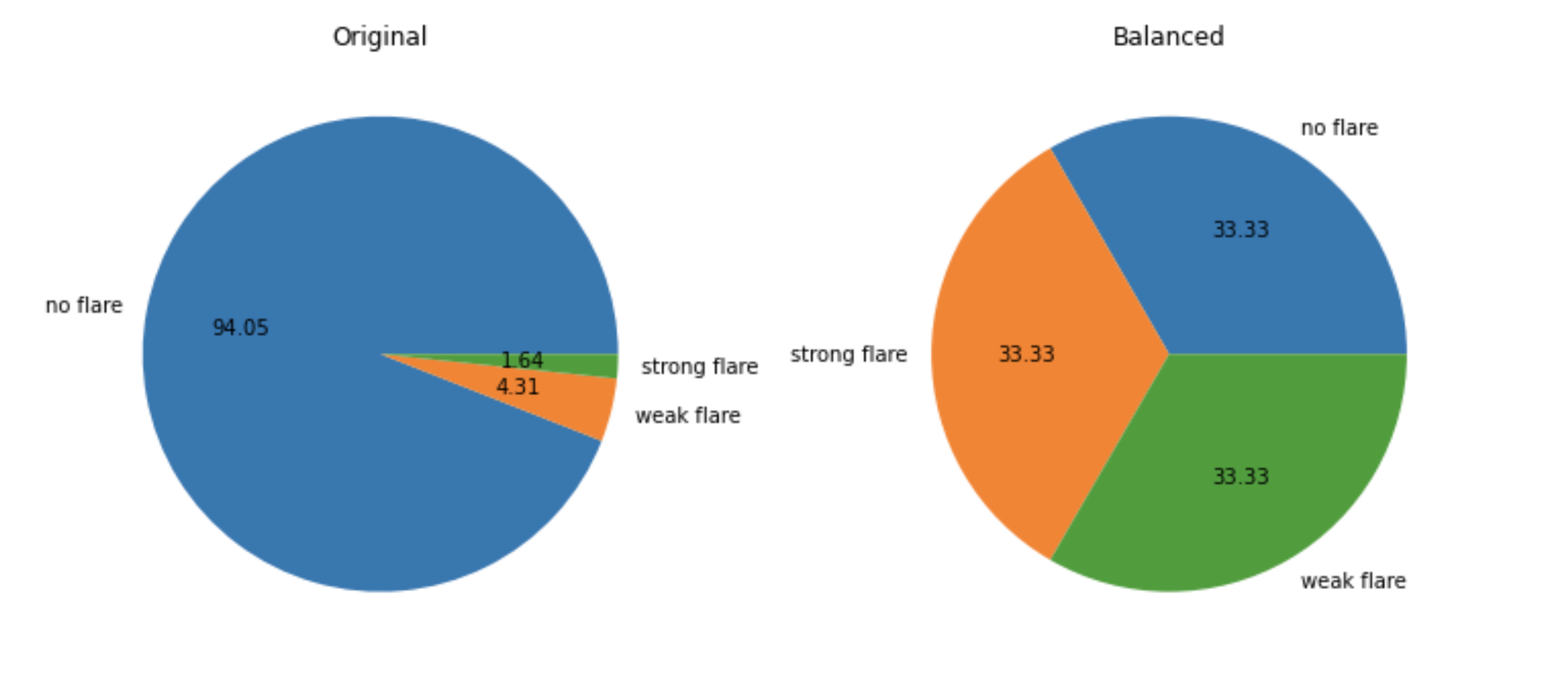}
\caption{Original solar flares data proportion (left) Balanced solar flares data (right)
\label{fig:undersampling}}
\end{figure}

 Under-sampling is a strategy that resolves the issue of imbalanced data by reducing the size of the over-represented class \citep{haibo_he_learning_2009, mosley_balanced_2013}. In our study, the dominant class was the no-flare class. Initially, we detected 28,415 flare events from the RHESSI dataset corresponding to solar spectra observations. Notably, the strong flare class exhibited the lowest representation, comprising only 467 instances. To preserve the model's ability to classify strong flares when under-sampling, we randomly selected 467 no-flare spectra and 467 weak flare spectra from the PCA 1000 components data using the \textit{make imbalance} function from the \textit{imbalanced-learn} module. The ratios of data before and after undersampling is represented by figure \ref{fig:undersampling}. For each class, there are 467 spectra, which makes the under-sampled set 1401 spectra.

\subsection{Testing Supervised-learning Models}
Since we divided the flares into three classes, we selected models that are best fit for multi-class training and compared their performance on two tasks: the average confusion matrices over 10 trials and categorical accuracy scores (see Section 4.1 Equation \ref{formula:CategoricalAccuracy} for definition of categorical accuracy scores). The models we selected are random forest, support vector machine with stochastic gradient descent, and C-Support Vector Classification, all implemented  with \textit{sklearn} \citep{scikit-learn}. \edit1{We selected these models because they are well-suited for supervised multi-label classification tasks and consistently outperformed other models designed for similar problems \citep{scikit-learn}.}

We evaluated the performance of each model using its confusion matrix averaged over 10 trials. \edit1{For each trial, we split the data into two sets: 80\% went into the training set, while 20\% went into the test set.} A confusion matrix is a structured arrangement used to visualize the effectiveness of an algorithm, often used within the context of supervised learning \citep{stehman_selecting_1997}. The matrix expands to an $n \times n$ format, where n is the number of categories or classes. In our case, we have a $3 \times 3$ matrix.  Each row represents the actual classes, while each column represents the predicted classes. Each entry in the matrix shows the number of observations from the actual class (row) that were predicted to be in a specific class (column) \citep{stehman_selecting_1997}. Therefore, visually, the brighter the diagonal of the heat map generated from the confusion matrix is, the more instances from each class are classified correctly.

We created the average confusion matrices by summing the confusion matrices obtained from individual trials entry by entry and then dividing them by the number of trials. We chose to run 10 trials because unlike reinforcement learning or deep learning models, there is less randomness in the training process for supervised learning, therefore it is not necessary to run the models for a large number of trials to get their average performance. 

\section{Results and Discussion} \label{results}

\subsection{Model Evaluation and Selection}
We trained four models and analyzed their performance to identify the most promising candidates for further tuning and optimization. Table \ref{table:modeldescriptions}  shows the descriptions of the models and their hyperparameters. Hyperparameters are parameters of the model that are not determined before the training process, which can be tuned to minimize the generalization error or underfitting \citep{probst_tunability_2018}. None of the models that we used in the study use the default hyperparameters from \edit1{\textit{sklearn}}. We selected these specific hyperparameters to optimize the training results based on experiments we ran previously.

\begin{table}
    \resizebox{9cm}{!}{
        \begin{tabular}{|c|p{5cm}|}
        \hline
        \textbf{Algorithm} & \textbf{Description of Hyperparameters} \\
        \hline
        Random Forest & 
        \begin{itemize}
        \item Number of Estimators: 100
        \end{itemize}
        \\
        \hline
        SGD Classifier & 
        \begin{itemize}
        \item Loss Function: Hinge
        \end{itemize}
        \\
        \hline
        C-Support Vector Classification & 
        \begin{itemize}
        \item Kernel: Polynomial
        \item Degree: 2
        \item C (Regularization Parameter): 1.0
        \end{itemize}
        \\
        \hline
        C-Support Vector Classification & 
        \begin{itemize}
        \item Kernel: Sigmoid
        \item C (Regularization Parameter): 10
        \item sigma: 0.0001
        \end{itemize}
        \\
        \hline 
        \end{tabular}
}
\caption{Description of Hyperparameters for Models}
\label{table:modeldescriptions}
\end{table}

\edit1{Kernels are mathematical functions to transform the non-linearly separable data into a higher-dimensional space, where the data can be separated by a linear equation \citep{scholkopf_primer_2004}. We tried 3 kernels: polynomial, sigmoid, and Radial Basis Function (RBF). RBF proved to have the best results on average. RBF is described by the following equation:
\begin{align}
    K(\mathbf{x}, \mathbf{x}') = \exp\left(-\frac{\|\mathbf{x} - \mathbf{x}'\|^2}{2\sigma^2}\right)
\end{align}
where $x$ and $x'$ are two distinct data samples from the dataset, and $\sigma$ is a free parameter \citep{scholkopf_primer_2004}.}

\begin{figure}[ht]
    \includegraphics[scale = 0.26]{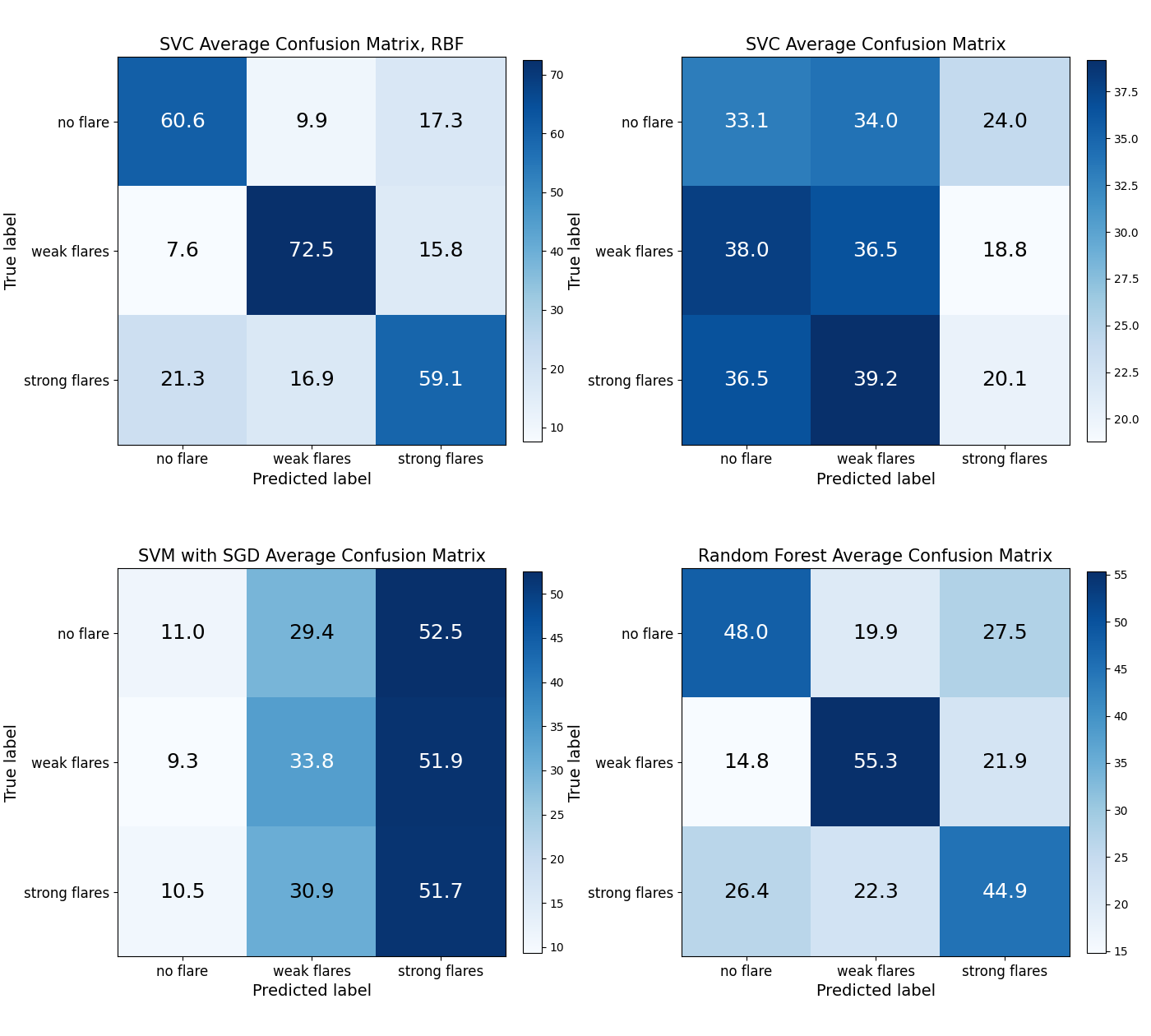}
    \caption{\edit1{Average Confusion Matrices for SVC with an RBF kernel (top left), SVC with Polynomial kernel (top right), SVM with Stochastic gradient descent optimizer (bottom left), Random Forest (bottom right)}}
    \label{fig:avgmatrices}
\end{figure}

Figure \ref{fig:avgmatrices} shows the average confusion matrices for the four models we trained. The average confusion matrices of SVC with an RBF kernel and Random Forest have the brightest diagonals, which shows that the models can predict most of the data points in each class correctly, especially when compared to the confusion matrices of SVC with polynomial kernel and SVM with stochastic gradient descent optimizer. This confirms that our models have learned to correctly detect the presence of solar flares events and classify some of them correctly using solar spectra.
We further evaluated the performance of Random Forest and SVC with an RBF kernel by comparing their accuracy scores and categorical accuracy scores. Accuracy score, or aggregated accuracy score as seen in the following Equation \ref{formula:AggregateAccuracy}, is a commonly used metric that calculates the overall accuracy of a classifier.
It measures the proportion of correctly predicted instances out of the total instances in the dataset \citep{mosley_balanced_2013}. \\


\begin{sizeddisplay}{\scriptsize}
\begin{align} \label{formula:AggregateAccuracy}
    \text{Aggregate Accuracy} = \frac{\text{Number of Correct Predictions}}{\text{Total Number of Predictions}}
\end{align}
\end{sizeddisplay}
Since aggregate accuracy is ``blind" to specific classes, we introduced another metric called the categorical accuracy score to evaluate the performance of the model for each class. Equation \ref{formula:CategoricalAccuracy} shows the formula for the categorical accuracy score, which is a variant of the aggregate accuracy score specifically designed for multi-class classification problems, like the problem in this study where there are three classes instead of only two \citep{mosley_balanced_2013}.
It calculates the accuracy considering each class separately. In other words, it calculates the proportion of correctly predicted instances for each class out of the total instances belonging to that class.\\

\begin{sizeddisplay}{\scriptsize}
\begin{align}\label{formula:CategoricalAccuracy}
\text{Categorical Accuracy} = \frac{\text{Number of Correct Predictions in the Class}}{\text{Total Number of Instances in the Class}}
\end{align}
\end{sizeddisplay}

Other than accuracy scores, we also used metrics like Precision and Recall class-wise to further evaluate the performance of each model. The formulas for Precision and Recall are \citep{kelleher_fundamentals_2015}:

\begin{sizeddisplay}{\scriptsize}
\begin{align} \label{formula:precision}
    \text{Precision}_{\text{Class A}} = \frac{\text{TP }_\text{Class A}}{ \text{TP}_\text{Class A } + \text{FP}_\text{Class A}}
\end{align}
\end{sizeddisplay}

\begin{sizeddisplay}{\scriptsize}
\begin{align}
\label{formula:recall}
    \text{Recall}_{\text{Class A}} = \frac{\text{TP }_\text{Class A}}{ \text{TP}_\text{Class A } + \text{FN}_\text{Class A}}
\end{align}
\end{sizeddisplay}
\noindent

where ``TP'' is True Positives, the number of instances that were positive in the dataset and were correctly classified as positive by the model, ``FN'' is False Negatives, the number of instances the model failed to identify as positive when they actually were, and ``FP'' represents False Positives, the number of instances the model mistakenly classified as positive when they were actually negative.
As seen in Equation \ref{formula:precision}, precision for a given class in multi-class classification is the fraction of instances correctly classified as belonging to a specific class out of all instances the model predicted to belong to that class \citep{kelleher_fundamentals_2015}. In other words, precision tells you the ratio of true positives to all instances that the model predicted as positive. Precision measures the accuracy of all positive predictions. \edit1{Recall for a given class is the ratio between correctly classified positive instances and all positive instances in the data \citep{kelleher_fundamentals_2015}. It measures whether all positive instances are predicted correctly.}

Another evaluation metric we used was the F1 score, which evaluates the performance of a model using both precision and recall. Equation \ref{eqn:F1} shows the formula for F1 score. 
\begin{align}\label{eqn:F1}
    F_1 = \frac{2}{\text{Recall}^{-1}+\text{Precision}^{-1}}
\end{align}
A high F1 score indicates a balanced performance, meaning that the model is both accurate in predicting positive instances and is able to classify most of the positive instances correctly \citep{article}.

\subsection{SVC with an RBF Kernel}
\begin{figure}
    \hspace*{-0.6cm} 
    \includegraphics[scale = 0.4]{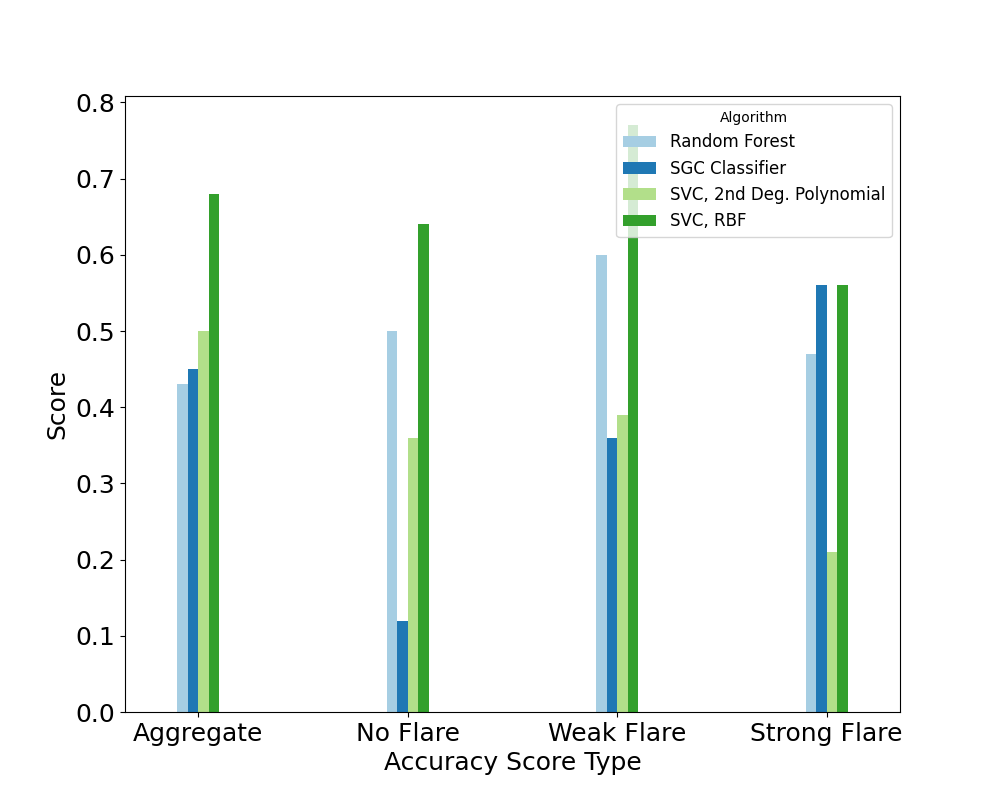}
    \caption{Categorical and aggregated accuracy score of each algorithm}
    \label{fig:accuracyscores}
\end{figure}

Figure \ref{fig:accuracyscores} shows that in our preliminary experiments with different models SVC with an RBF kernel had the best performance in both aggregate and categorical accuracy scores for all three classes. The average aggregate accuracy score over 10 trials is 0.68. The average categorical accuracy scores for each of the classes\edit1{ -- no flare, weak flares, and strong flares -- }are 0.64, 0.77, and 0.56, respectively. 

\begin{figure}
    \hspace*{-0.6cm} 
    \includegraphics[scale = 0.4]{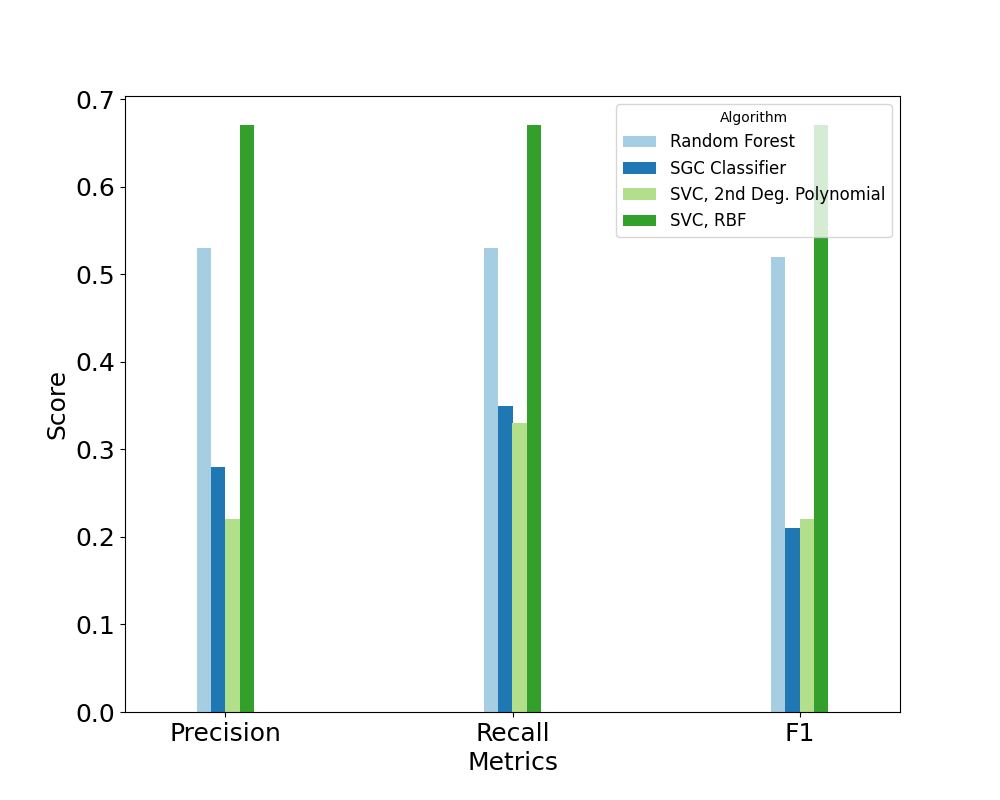}
    \caption{Average precision, recall, and F1 scores of each algorithm}
    \label{fig:PrecisionsandRecalls}
\end{figure}

Figure \ref{fig:PrecisionsandRecalls} shows the average Precision, Recall, and F1 scores of each model. The Random Forest model demonstrates a balanced performance with both Precision and Recall at around 0.53. This suggests that it maintains a good equilibrium between correctly identifying positive cases and minimizing false positives. The F1 score of 0.52 further reinforces this balance, making it a solid choice when a trade-off between Precision and Recall is required. The SGD Classifier, on the other hand, exhibits lower Precision (0.28) and Recall (0.35) compared to the Random Forest. This indicates that it might struggle with both correctly classifying positive instances and minimizing false positives. The F1 score of 0.21 underscores this performance gap, implying that this model may not be the best option in situations where Precision and Recall are critical. The SVC with a 2nd-degree polynomial kernel shows similarly suboptimal performance with a low Precision of 0.22 and Recall of 0.33. Its F1 score of 0.22 suggests that it doesn't excel at either precision or recall. This model appears to underperform in comparison to the Random Forest. The SVC with an RBF kernel has the highest Precision, Recall, and F1 score, all at 0.67. This model performs well in both correctly classifying positive instances and minimizing false positives compared to the other models. Based on these results, We decided to further optimize the SVC model with an RBF kernel.


\begin{figure}[h] 
\centering
\includegraphics[scale = 0.55]{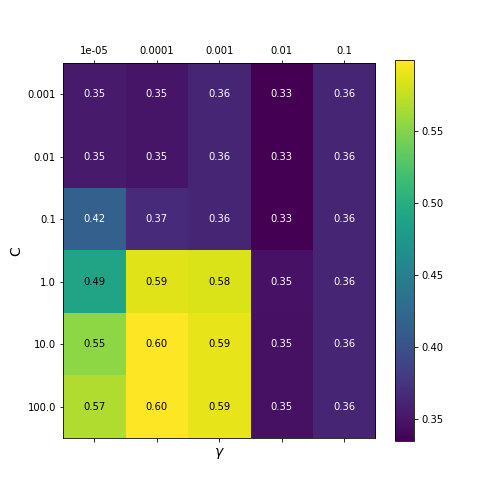}
\caption{Grid Search, aggregate accuracy and corresponding C and $\gamma$ values of SVC with an RBF kernel}
\label{fig:rbf_grid}
\end{figure}

The \textit{SVC} function from \textit{scikit-learn} with RBF kernel has two hyper-parameters, C and $\gamma$, that allow us to find the right value of variance that optimizes the model performance with the RBF Kernel function \citep{scikit-learn}. C value is the regularizer parameter that can be manipulated to control the extent of overfitting. A high C value may lead to overfitting, and a low C value may lead to underfitting, where the model is too generalized to identify the pattern in training data. $\gamma$ defines how far the influence of a single training example reaches. A high $\gamma$ value suggests ``close" and may lead to overfitting because it requires the data points to be close to group them; a low $\gamma$ value suggests ``far" and may lead to under-fitting \citep{scikit-learn}. We applied \textit{GridSearh CV} from \textit{sklearn} to determine which combination of C value and $\gamma$ value result in the best aggregate accuracy score.

To determine good values of the hyperparameters, it is important to search on the right scale. \citep{chih-wei_hsu_practical_2016} We decided to test a range of C values, 0.001, 0.01, 0.1, 1.0, 10.0, and 100.0, aiming to encompass a broad spectrum, since the most common range of C values to test is \edit1{1 -- 10} \citep{chih-wei_hsu_practical_2016}. Initially, our empirical testing focused on the common range of 0 to 1.0 for C. Notably, we observed a substantial increase in accuracy scores as C surpassed 1, particularly with $\gamma$ values of 0.00001, 0.0001, and 0.001. However, for C values higher than 10.0, we detected no significant accuracy improvement; instead, there was a slight decrease. Moreover, C values exceeding 100 exhibited a higher likelihood of leading to overfitting. Consequently, we decided to stop further testing C values higher than 100.

Similarly, we tried different $\gamma$ values to find one that balances the variance and the bias of our model. The $\gamma$ value corresponds to the margin of the kernel function in the higher-dimensional space that training data were projected onto. In this higher dimensional space, we partition the training data into 3 sections, corresponding to the number of classes in our multi-class classification problem. The boundaries at the intersection of these 3 spaces are called the decision boundaries \citep{kelleher_fundamentals_2015}. A high $\gamma$ value means only the closest points to the decision boundary will carry the weight leading to a smoother boundary, which likely results in over-fitting. Whereas a low $\gamma$ value corresponds to a larger margin that contains more data points, which leads to under-fitting. We tested a range of $\gamma$ values to see their effects on the training results and whether decreasing the $\gamma$ value would lead to high-accuracy model performance. The values chosen were 0.00001, 0.0001, 0.001, 0.01, and 0.1. We chose this range because, theoretically, choosing exponentially growing sequences of C and $\gamma$ values is more efficient when determining good parameters using grid search \citep{chih-wei_hsu_practical_2016}. In practice, it is good to try a $\gamma$ value that is $6/k$, where k is the number of input data samples \citep{Chapelle2005SemiSupervisedCB}. Here we have 1401 data points, which theoretically makes $6/1401 = 0.002$ an ideal $\gamma$ value. Nonetheless, using a moderately coarse grid helps to identify the optimal region within it. Then, we can empirically determine which region on the coarse grid results in better performance. This is why we chose a slightly wider range of C and $\gamma$ values compared to common practice.

As seen in Figure \ref{fig:rbf_grid}, when $C = 10.0$ and $\gamma = 0.0001$, the SVC model with an RBF kernel achieves an aggregate accuracy score of 0.60, which is the highest out of all tested combinations. We identified that the model performance is optimal when C is between 10.0 and 100.0, and when $\gamma$ is between 0.0001 and 0.001.

\begin{figure} 
\hspace*{-0.5cm} 
\includegraphics[scale = 0.4]{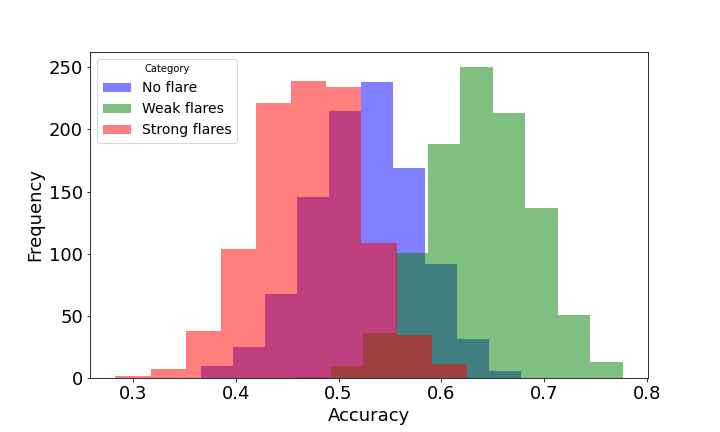}
\caption{Accuracy score distribution across 1000 trials for no flare, weak flares, and strong flares}
\label{fig:rbf_grid}
\end{figure}

\section{Conclusions}


Based on the results presented above, we concluded that it is possible to detect and classify solar flares in optical high-resolution spectra using supervised learning algorithms. We used SVC with an RBF kernel to categorize solar flares into three classes. The model exhibited an overall accuracy score of 0.65, showcasing its ability to distinguish among these distinct flare categories. A blind classification algorithm would have an accuracy score of 0.33, so our algorithm is a significant improvement. Nonetheless, there are some apparent limitations to our findings. 

One limitation is the model's comparatively low accuracy in classifying the ``strong flare" category, as evident from a categorical accuracy of 0.56. A possible strategy for improvement is supplementing the training data with more actual data on strong flares; increasing the model's exposure to this class should enhance its ability to classify strong flares accurately. 

Another limitation is the overall performance of the SVC model with the RBF kernel. The aggregate accuracy shows that more than half of the time the model can classify most data points correctly, but there is still room for performance enhancement. The parameter choices, such as C (set at 10) and $\gamma$ (0.0001), may not be optimally configured, necessitating a more comprehensive exploration of hyperparameter settings. To further improve the overall accuracy of our current SVC model, a multifaceted approach can be adopted. First, fine-tuning the hyperparameters, such as the regularization parameter (C) and the kernel-specific parameter ($\gamma$), through a more extensive grid search can lead to an optimized model configuration. This would enable us to strike the right balance between model complexity and generalization. 

Also, while there are documented time differences between these regions, as supported by relevant literature, these differences are not consistently predictable. Given the variability and our lack of precise knowledge about these time discrepancies, our investigation does not explicitly account for them. Additionally, solar flares, which are the focus of our study, persist over extended periods rather than occurring as instantaneous events. Our approach assumes the effectiveness of our current method, and the performance of our final machine learning model supports this assumption, showing satisfactory results.

Moreover, our findings reveal implications for the characteristics of the underlying data. Notably, the RBF kernel demonstrated significant performance in classifying weak flares, both before and after hyperparameter tuning. The model achieves an average categorical accuracy score of 0.77 before tuning, and an average categorical accuracy score of 0.80 after tuning. This suggests that weak flares may exhibit distinct and non-linear patterns effectively captured by the RBF kernel. In other words, the data may not be homogeneous, and the classes may have varying degrees of complexity.
One possible explanation could be that all of the ``weak flares" are similar to each other, while the strong flares represent a larger range of flare energies. Therefore, when our model is tested on the testing data, it might be less accurate when predicting strong flares compared to weak flares.  

In future work, one could explore other learning algorithms to determine if they achieve better performance. Ensemble learning methods, such as Random Forest and Gradient Boosting, harness the collective power of multiple models and potentially improve overall classification accuracy. Also, given the non-linear nature of the underlying data, as evidenced by the performance of the RBF, it makes sense to consider the potential of employing deep learning techniques as a next step. Deep learning could potentially address the local cluster patterns within our high-dimensional data due to its capability to uncover hidden structures and nuances that may elude traditional machine-learning models. It is possible that a neural network with many neurons can preserve and identify the complex patterns in strong flare data that supervised learning models struggle to capture. 

Our work reported here represents an initial investigation aimed at automating the detection and classification of flares in high-resolution solar spectra. While we have achieved the development of a model capable of classifying solar flares within this context, future efforts should focus on enhancing the prediction accuracy and exploring the potential of alternative models, including deep learning approaches, to further refine the classification capabilities.

The longer-term vision is to develop a robust framework for detecting and categorizing stellar flares not just in our solar system, but also in the broader context of exoplanetary systems. The latter would enable more accurate corrections for stellar contamination. Extending such an approach to the host stars of exoplanets involves adapting the SVC models to cater to the particular characteristics of these stars' spectra. Since the spectral signatures and flare activities of exoplanet host stars might differ from those of the Sun, the model would require recalibration and retraining with relevant datasets. The recalibration would involve adjusting the model to recognize flare signatures in different stellar environments, taking into account factors such as the star's size, age, and magnetic activity. 

\section{Acknowledgements}
We thank the anonymous referee for their helpful comments.  NH acknowledges support from the Nexus Scholars prorgam in the College of Arts and Sciences at Cornell University.




\section{Code Availablity}
To access the code used in this project, please use this link: \url{https://github.com/nicolehao34/solar_flares_classification}

\newpage 

\vspace{1cm}
\bibliographystyle{aasjournal}


\end{document}